\documentclass[style=bcnpostgrad,entry,preprint]{proceedings}
\renewcommand{\conferencetitle}{Barcelona Postgrad Encounters on Fundamental Physics}
\usepackage[utf8]{inputenc}

\begin{document}

\author{Martin Kr\v{s}\v{s}\'ak}
\email{krssak@physik.uni-bielefeld.de}
\affiliation{Faculty of Physics, University of Bielefeld, \\ D-33501 Bielefeld, Germany}

\title{Viscosity Correlators in Improved Holographic QCD} 

\abstract{We study a bottom-up holographic model of large-$N_c$ Yang-Mills theory, in which conformal invariance is broken through the introduction of a dilaton potential on the gravity side.  We use the model to calculate the spectral densities of the shear and bulk channels at finite temperature. In the shear channel, we compare our results to those derived in strongly coupled ${\mathcal N}=4$ Super Yang-Mills theory as well as in weakly coupled ordinary Yang-Mills. In the bulk channel, where the conformal result is trivial, we make comparisons with both perturbative and lattice QCD. In both channels, we pay particular attention into the effects of conformal invariance breaking in the IHQCD model.} 

\acknowledgements{The author would like to thank Keijo Kajantie and Aleksi Vuorinen for collaboration, and in addition Yan Zhu for sharing her perturbative QCD results in a useful form as well as Harvey Meyer for providing his lattice data. The work was supported by the DFG graduate school \textit{Quantum Fields and Strongly Interacting Matter} as well as the Sofja Kovalevskaja programme of the Alexander von Hulboldt foundation. I would also like to express my gratitude to the organizers of this conference for hospitality and for the possibility to present this talk.}

\maketitle

\section{Introduction}
Heavy ion collision data from the RHIC \cite{Tannenbaum:2012ma} and LHC \cite{Muller:2012zq} experiments suggests that quark-gluon plasma near $T_c$ behaves as an almost ideal liquid with a  very low shear viscosity to entropy ratio, $\eta/ s \le 0.2$. Understanding this small value has turned out to be very difficult using standard field theory methods. Calculations in perturbative QCD suggest that this ratio should be $\eta/ s \approx 1$ \cite{Arnold:2003zc}, in direct contradiction with the experiments. Moreover, the shear viscosity is a real-time transport coefficient, so it is very difficult to calculate in lattice QCD. 

One of the most famous predictions resulting from the AdS/CFT conjecture has been the existence of a universal bound for the viscosity to entropy ratio $\eta/ s \geq 1/(4 \pi)$ \cite{Kovtun:2004de}, which no physical liquid is supposed to  violate. Since then, AdS/CFT methods have played an important role in increasing our understanding of strongly interacting systems, such as the matter produced in heavy ion collisions \cite{CasalderreySolana:2011us}. However, AdS/CFT is originally a duality between string theory and conformally invariant  supersymmetric Yang-Mills theory. As QCD is not conformal nor supersymmetric, it would be important to develop a holographic model with these features, which might be able to describe real-world QCD more closely. One such model is the Improved Holographic QCD (IHQCD) proposed by Elias Kiritsis, et.~al.~\cite{Gursoy:2007cb,Gursoy:2007er}, where conformal invariance is broken by the introduction of a nontrivial potential for the dilaton field, mimicking perturbative QCD in the UV region and giving the model a linear glueball spectrum in the IR, i.e.~making it confining.

In this talk, we present results published in the papers \cite{Kajantie:2011nx,Kajantie:2013gab}, where using IHQCD we have calculated the field theory correlators
\begin{eqnarray}
G_s^R(\omega,\mathbf{k}=0)&=&-i\int\! {\rm d}^4x\,  e^{i \omega t} \theta (t) \langle[
T_{12}(t,\vec{x}),T_{12}(0,0) ]\rangle \label{Gsdef}\, \\
G_b^R(\omega,\mathbf{k}=0)&=&-i\int\! {\rm d}^4x\,  e^{i \omega t} \theta (t) \langle[\frac{1}{3}
T_{ii}(t,\vec{x}),\frac{1}{3}T_{jj}(0,0) ]\rangle \label{Gbdef} \, ,
\end{eqnarray}
where $T_{12}$ and $T_{ii}$ are the shear and bulk components of the energy-momentum tensor, respectively. Of our primary interest are the spectral densities, which are defined as 
\begin{equation}
\rho_{s,b} (\omega,T)=\mathop{\mbox{Im}} G_{s,b}^R(\omega,\mathbf{k}=0)\, .
\end{equation} 
Using the Kubo formulas, we can furthermore formulate the relations 
\begin{eqnarray}
\eta &=&\lim_{\omega\to 0}\frac{\rho_s(\omega,T)}{\omega}\, ,\label{eta} \\
\zeta &=&\lim_{\omega\to 0}\frac{\rho_b(\omega,T)}{\omega}\, \label{zeta}
\end{eqnarray}
between the spectral functions and the corresponding transport coefficients, the shear ($\eta$) and bulk ($\zeta$) viscosities of the theory.

As IHQCD is a two-derivative model, we expect the shear viscosity to entropy ratio to have the universal value $\eta/s=1/(4 \pi)$, as proven in general in \cite{Iqbal:2008by}. This is in fact one of our reasons to focus on the properties of the spectral densities, as they allow us to distinguish between strongly coupled ${\mathcal N}=4$ SYM and the field theory dual to IHQCD. In the shear channel, we furthermore have available a recent pertubative result for the Yang-Mills spectral density \cite{Zhu:2012be}, which naturally is not applicable in the region of low frequencies, but provides us with an important consistency check at higher frequencies. In the bulk channel, the SYM spectral density (and bulk viscosity) on the other hand vanish due to conformal invariance, so a comparison of our IHQCD results with the perturbative result of \cite{Laine:2011xm} and the lattice data of \cite{Meyer:2010ii} (for Euclidean imaginary time correlators) is of utmost interest.

\section{Improved Holographic QCD}
Improved Holographic QCD is a five-dimensional bottom-up holographic model, where on the gravitational side one starts with a gravity+dilaton system and introduces a potential for the dilaton field to describe some crucial features of conformal symmetry breaking \cite{Gursoy:2007cb,Gursoy:2007er}. The action reads
\begin{equation}
S=\frac{1}{16 \pi G_5}\int {\rm d}^5x\sqrt{-g} \left[R -\frac{4}{3}\left(\frac{\partial \lambda}{\lambda}\right)^2 +V(\lambda) \right]\, ,  \label{action}
\end{equation}
where we have used $\lambda=e^\phi$, with $\phi$ denoting a dilaton field. The form of the potential $V(\lambda)$ is found by matching the holographic beta function to the 2-loop perturbative one (in large-$N_c$ Yang-Mills theory) and by requiring the model to possess a linear glueball spectrum. This can be achieved, for example, through the potential \cite{Jarvinen:2011qe}
\begin{equation}
V(\lambda)=\frac{12}{\mathcal{L}^2}\left[
1+\frac{88}{27}\lambda + \frac{4619}{729}\lambda^2\frac{ \sqrt{1+\ln(1+\lambda)}}{(1+\lambda)^{2/3}}
\right]\, . \label{V}
\end{equation}
Note that this is the potential we have used in our most recent publication \cite{Kajantie:2013gab}, while in \cite{Kajantie:2011nx} a slightly different choice, consistent with the original potential of \cite{Gursoy:2007cb,Gursoy:2007er}, was made. For  possible issues in the derivation of the potential we would like to point on  \cite{Veschgini:2010ws,Megias:2010tj}. 

A background metric consistent with the above potential reads
\begin{equation}
ds^2=b^2(z)\left[-f(z)dt^2+d{\bf x}^2 +{dz^2\over f(z)}\right]\, ,  \label{metric}
\end{equation}
where the functions $f(z)$, $b(z)$ and $\lambda(z)$ are determined from the Einstein equations (\ref{action})
\begin{eqnarray}
\dot{W}&=& 4 b W^2 -\frac{1}{f}(W\dot{f} +\frac{1}{3}b V ),  \label{Einstein1}\\
\dot{b} &=&-b^2 W\,, \label{Einstein2}\\
\dot{\lambda} &=&\frac{3}{2}\lambda \sqrt{b \dot{W}}\,,\label{Einstein3}\\
\ddot{f} &=& 3\dot{f}b W\,.\label{Einstein4}
\end{eqnarray}
An additional constraint is that the space-time be asymptotically AdS, i.e.~that close to the boundary
\begin{equation}
b(z)\to\frac{\mathcal{L}}{z}, \ \ \  \ \ \ \ z \to 0.
\end{equation}
where $\mathcal{L}$ is the curvature radius of the AdS space-time, which without loss of generality we can set to $\mathcal{L}=1$. Using eqs.~(\ref{Einstein1})-(\ref{Einstein4}), we can next determine the thermodynamic properties of the system \cite{Gursoy:2008za,Alanen:2011hh}. Matching these results to the  corresponding quantities in large-$N_c$ Yang-Mills theory, we are able to finally fix the value of the gravitational constant 
\begin{equation}
\frac{\mathcal{L}^3}{4\pi G_5}=\frac{4 N_c^2}{45 \pi^2}\, .
\label{ellcube}
\end{equation}

To determine correlators of the energy-momentum tensor on the field theory side, we follow \cite{ Gubser:2008sz,Gursoy:2009kk} and introduce perturbations around the background metric (\ref{metric}),
\begin{eqnarray}
g_{00} = b^2 f \left(1+\epsilon H_{00}\right), 
g_{11} = b^2 \left(1+\epsilon H_{11}\right), 
g_{55} =\frac{b^2}{f} \left(1+\epsilon H_{55}\right),
g_{12}=\epsilon b^2 H_{12}\, .
 \label{bulkpert}
\end{eqnarray}
Here, $\epsilon$ is a power counting parameter, and the fluctuation $H_{12}$ corresponds to the $T_{12}$ operator, while the fluctuation $H_{11}$ is dual to $\frac{1}{3}T_{ii}$. Expanding next the Einstein equations to the first order in $\epsilon$, we obtain fluctuation equations for the perturbations
\begin{eqnarray}
&&\ddot H_{12}+ {d\over dz}\log(b^3f)\dot H_{12}+
{\omega^2\over f^2} H_{12}=0 \,, \label{fluctshear} \\
&&\ddot H_{11}+ {d\over dz}\log(b^3fX^2)\dot H_{11}+
\biggl({\omega^2\over f^2} -{\dot f\,\dot X\over fX}\biggr)H_{11}=0\,,\label{fluctbulk}
\end{eqnarray}
where 
\begin{equation}
X(\lambda) \equiv  \frac{\dot{\lambda}}{3\lambda \dot{b}/b}. \label{Xdef}
\end{equation}
These equations are to be solved using infalling boundary conditions at the horizon \cite{Kajantie:2013gab}.

Having obtained the solutions to the above fluctuation equations, the spectral functions are available using standard results \cite{Gubser:2008sz},
\begin{eqnarray}
\rho_s(\omega,T)&=&\frac{s(T)}{4 \pi}\frac{\omega}{|H_{12}(z\rightarrow 0)|^2}\,, \label{rhos1} \\
\rho_b(\omega,T)&=&6X_h^2 \frac{s(T)}{4 \pi}\frac{\omega}{|H_{11}(z\rightarrow 0)|^2}\, .\label{rhob1}
\end{eqnarray}
In the latter bulk channel result, $X_h$ denotes the quantity of eq.~(\ref{Xdef}) evaluated at the horizon. When comparing the holographic results to those of lattice QCD, another useful observable turns out to be the directly measurable Euclidean imaginary time correlator. This quantity is obtained from the spectral density through the relation
\begin{equation}
G(\tau,T)=\int_0^\infty\frac{d\omega}{\pi}\rho(\omega,T)\frac{\cosh\left[\left(
\frac{\beta}{2} -\tau\right) \pi \omega\right]}{\sinh\left( \frac{\beta}{2}\omega\right)}\, ,\quad \beta\equiv 1/T\, .
\label{Gtau}
\end{equation}

\section{Results}
Below, we present our results for the spectral densities, covering the cases of the shear and bulk channels separately. 

\subsection{Shear Channel}

\begin{figure}[t]
\centering
\includegraphics[width=0.48\textwidth]{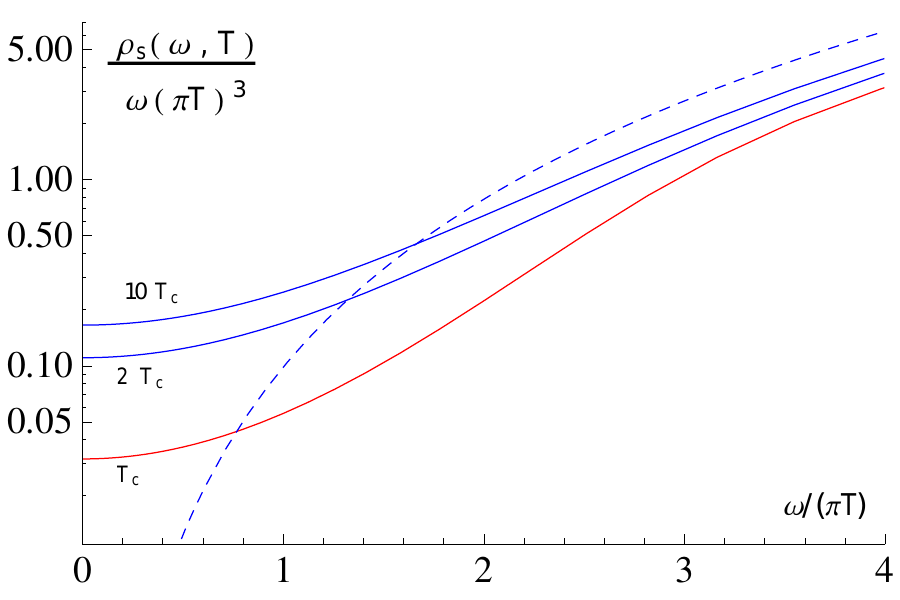}$\;\;\;$ \includegraphics[width=0.48\textwidth]{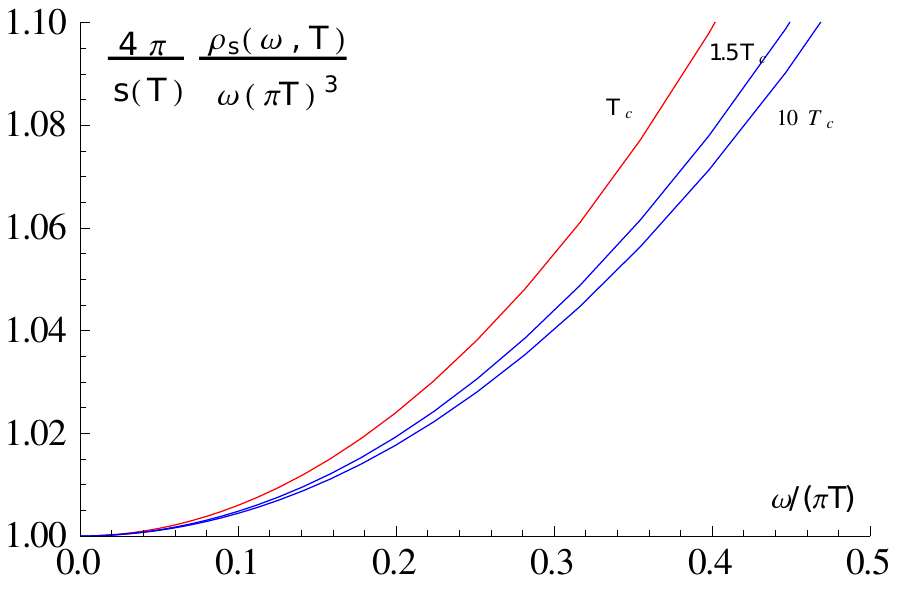}
\caption{\small Left: the ratio of the IHQCD  spectral density  $\rho_s(\omega)$ and frequency $\omega$ in the shear channel, displayed for three different temperatures in units of $\mathcal{L}^3/(4\pi G_5) $. The dashed curve corresponds to the asymptotic limit of $\pi\omega^3/32$. 
Right: the ratio of the spectral density and frequency normalized by the entropy. Using the Kubo formula of eq.~(\ref{eta}), we observe that the shear viscosity over entropy obtains the universal value of $1/(4\pi)$.     }
\label{figshear}
\end{figure}

The ase of the shear channel was considered already in \cite{Kajantie:2011nx}, which, unlike \cite{Kajantie:2013gab}, however did not make use of the potential of eq.~(\ref{V}). Thus, we will here concentrate on the results for the latter reference, showing in fig.~\ref{figshear} (left) the ratio of the spectral density and $\omega$ for various temperatures.  We observe a behavior similar to that displayed in fig.~3 of \cite{Kajantie:2011nx}; however, due to the choice of potential the logarithmic convergence of our result towards its large-$\omega$ limit ($\rho_{as}=\pi \omega^4/32$ in units where $\mathcal{L}^3/(4\pi G_5)$ has been set to unity) is somewhat slower.  Analogous calculations in $\mathcal{N}=4$ SYM theory (cf.~\cite{Kajantie:2010nx}) show that for temperatures close to the critical one, a significant difference between the IHQCD and SYM results arises, which however rapidly disappears at higher temperatures.

In fig.~\ref{figshear} (right), we next plot $\rho_s/\omega$ normalized by the entropy, from where we can read off the value of the shear viscosity over entropy ratio can as the intercept of the curves  at $\omega\to0$. As a consequence of  IHQCD being a two-derivative model, we find the universal prediction of $\eta/s=1/(4\pi)$ to hold at all temperatures considered \cite{Iqbal:2008by}. 

\begin{figure}[t]
\centering
\includegraphics[width=0.48\textwidth]{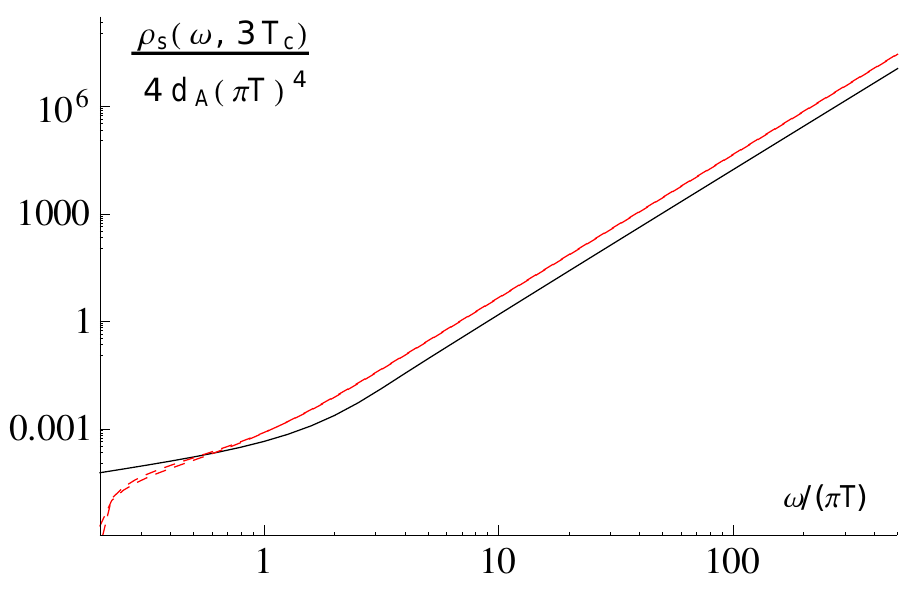}$\;\;\;$ \includegraphics[width=0.48\textwidth]{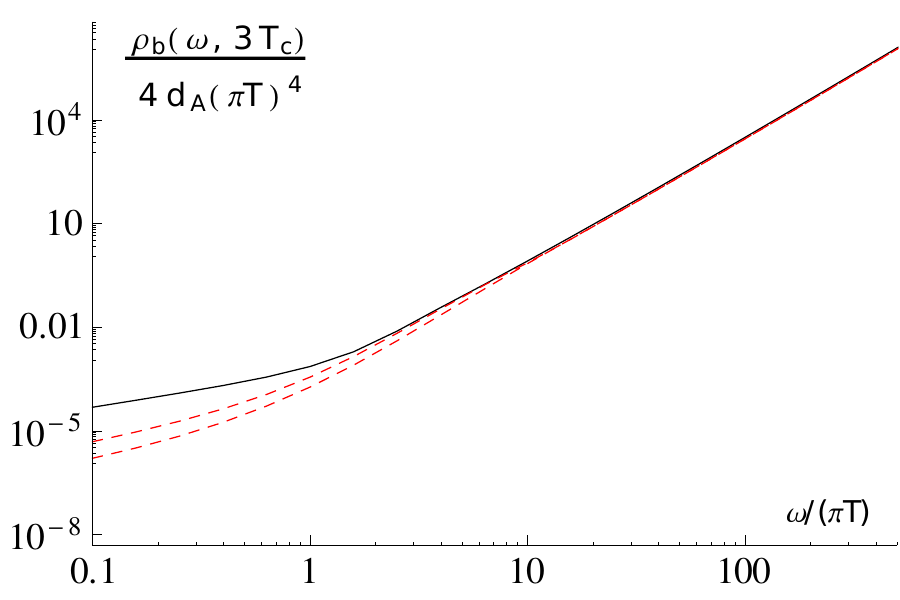}
\caption{\small Spectral densities in the shear (left) and bulk (right) channels for $T=3 T_c$, normalized by $4d_A=4(N_c^2-1)$. The black curves correspond to the IHQCD and the dashed dashed curves to the 2-loop perturbative QCD results \cite{Zhu:2012be,Laine:2011xm}.}
\label{figlarge}
\end{figure}

Despite the limitations of perturbative QCD in the region of low frequencies, at high $\omega$ the physical behavior of spectral densities is expected to reduce to its predictions due to asymptotic freedom \cite{CaronHuot:2009ns,Schroder:2011ht,Zhu:2012be}. To this end, in fig.~\ref{figlarge} (left), we plot both the perturbative (dashed red curve) and IHQCD (black curve) results for the shear spectral density, finding that in the UV limit, the IHQCD spectral function has the same functional behaviour as the perturbative one up to some overall normalization constant. As explained in detail in \cite{Kajantie:2013gab}, this proportionality factor can be identified as $4/9$ by analytically determining the asymptotic limits of both the perturbative and holographic spectral densities.    

\subsection{Bulk Channel}
\begin{figure}[t]
\centering
\includegraphics[width=0.48\textwidth]{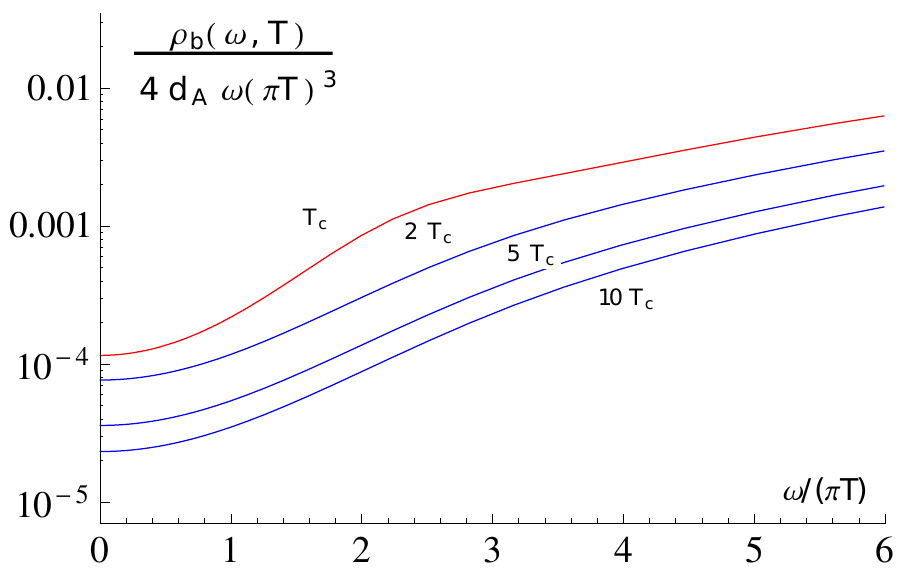}$\;\;\;$ \includegraphics[width=0.48\textwidth]{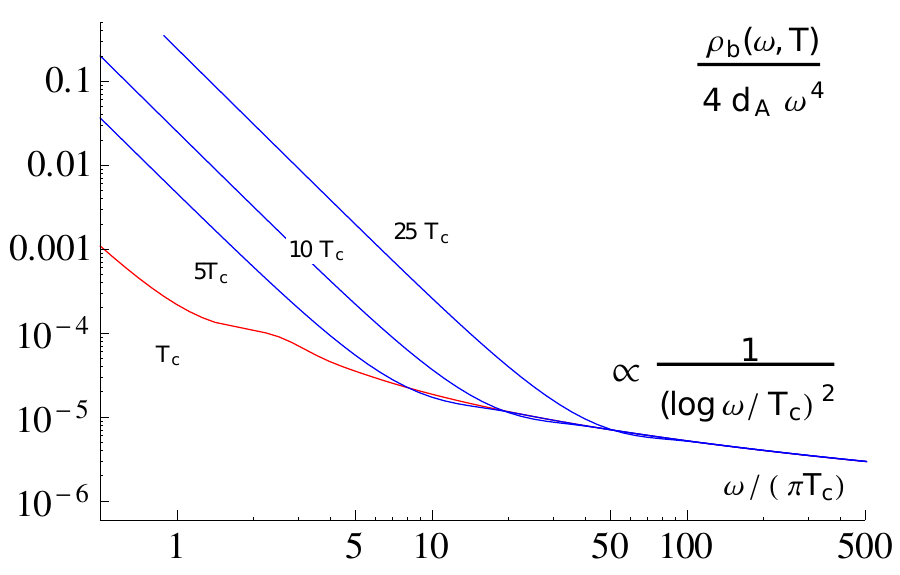}
\caption{\small Left:  the ratio of the IHQCD bulk spectral density  $\rho_b(\omega)$ and $\omega$, normalized by $4 d_A$. Right: the ratio $\rho_b/\omega^4$, plotted as a function of $\omega/(\pi T_c)$ for multiple temperatures, normalized by $4 d_A$. For large values of $\omega$, this ratio reduces to the  temperature independent limit $1/(\log\omega/T_c)^2$.}
\label{figbulk}
\end{figure}

Next, we move on to present the results of our recent correlator calculations in the bulk channel \cite{Kajantie:2013gab}. In fig.~\ref{figbulk} (left), we first plot $\rho_b(\omega)/\omega$ for different temperatures, normalized by the factor $4d_A$. Using the Kubo formula (\ref{zeta}), the bulk viscosity can again be read off as the value of the curves in the limit $\omega\to 0$. From these results, we see that the bulk viscosity decreases with increasing temperature, ultimately approaching the vanishing conformal limit. In fig.~\ref{figlarge} (right), we next find that in the region of large $\omega$, the IHQCD spectral function reduces to the perturbative QCD prediction. This reduction to is even more apparent when we study ratio the  $\rho_b/\omega^4$  plotted in fig.~\ref{figbulk} (right), where we see that this ratio quickly approaches the temperature independent limit $\propto 1/(\log\omega/T_c)^2$. This form in fact coincides with the $\omega$-dependence of the field theory gauge coupling $g^4$, to which the leading order perturbative result is proportional to \cite{Laine:2011xm,Laine:2010tc}.

\begin{figure}[t]
\centering
\includegraphics[width=0.48\textwidth]{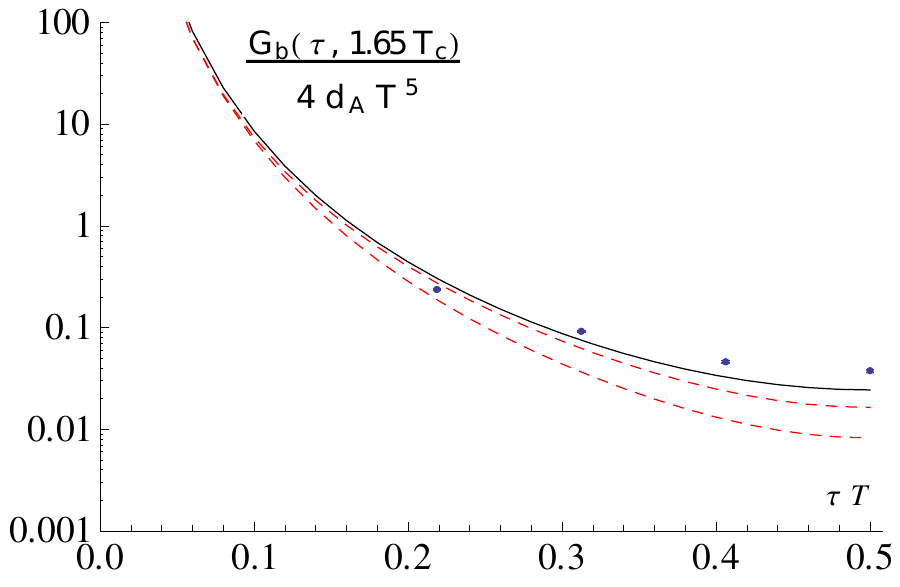}$\;\;\;$ \includegraphics[width=0.48\textwidth]{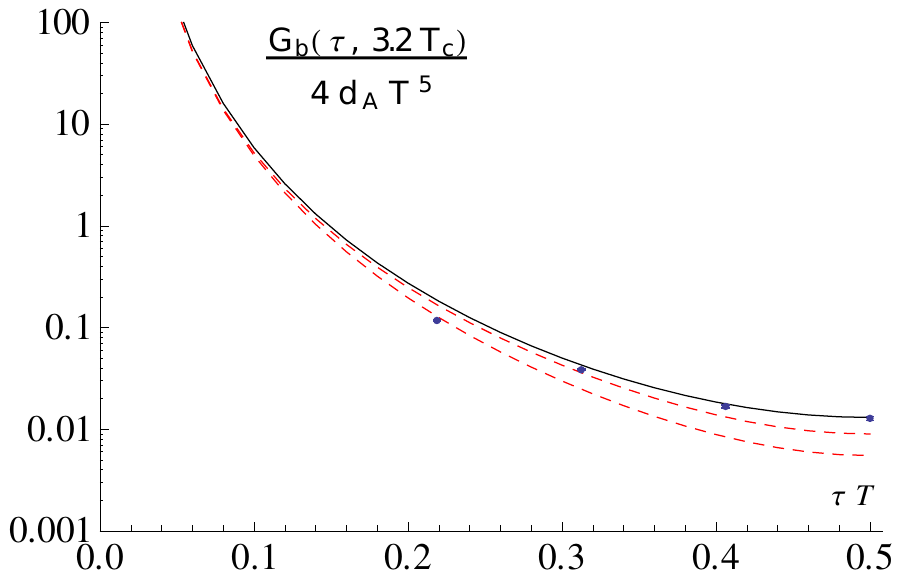}
\caption{\small The imaginary time correlators of the bulk channel, computed for two different temperatures in IHQCD (black curves) and pQCD (red dashed curves) \cite{Laine:2011xm} and compared with the lattice data (blue points) of \cite{Meyer:2010ii}.}
\label{figimaginary}
\end{figure}

Finally, lattice simulations of course play an important role in the study of strongly interacting phenomena, as they are the only fundamentally nonperturbative first principles method available. However, due to technical reasons only Euclidean correlators are possible to measure on the lattice, and to this end we next specialize to the case of imaginary time correlators, cf.~eq.~(\ref{Gtau}). With this result, it is possible to calculate Euclidean correlators in the both IHQCD and pQCD, and furthermore compare the results with the lattice data of \cite{Meyer:2010ii}, see fig.~\ref{figimaginary}. We find that the holographic results seem to be in better agreement with the lattice data than the perturbative ones for all temperatures considered. 

\section{Conclusions}
In this talk, we have presented the results of a recent calculation of finite temperature correlation functions in both the shear and bulk channels of  strongly coupled large-$N_c$ Yang-Mills theory. After motivating our research and stressing the importance of using non-conformal holographic models to describe the physics of QCD, we briefly introduced the fundamental properties of the IHQCD model and showed how spectral densities can be obtained in this framework.  

Next, we compared our holographic results with corresponding calculations in strongly coupled $\mathcal{N}=4$ supersymmetric Yang-Mills theory \cite{Kajantie:2010nx}, perturbative Yang-Mills theory \cite{Zhu:2012be,Laine:2011xm} as well as lattice simulations in the same theory \cite{Meyer:2010ii}. In the shear channel, we observed a significant difference between the IHQCD and SYM results close to the critical temperature $T_c$, which however rapidly disappears at higher temperatures. We also  compared the IHQCD and perturbative spectral densities in the large-$\omega$ region, finding in both cases an asymptotic $\omega^4$ behavior, albeit with differing overall normalizations.

Similar methods were next used to calculate the IHQCD spectral density in the bulk channel, where we confirmed that with increasing temperature, the bulk viscosity approaches the vanishing result known to occur in conformal theories. Subsequently, we compared our holographic calculation with results from perturbative Yang-Mills theory, finding impressive agreement in the UV (large-$\omega$)limt, including a perfectly matching overall normalization. We find this observation somewhat puzzling, considering the qualitatively different conclusion reached in the shear channel.

Finally, in the bulk channel, we calculated imaginary time correlators using both the IHQCD and perturbative QCD spectral functions, and compared the results with available lattice data. We conclude that the lattice data seems to favour our holographic results over perturbative one for all temperatures considered.   

\end{document}